\documentclass[aps,showkeys,secnumarabic,twocolumn,nobalancelastpage,amsmath,amssymb,nofootinbib,floatfix,fleqn]{revtex4}
\usepackage{amsfonts}
\usepackage{amsmath}
\usepackage{amssymb}
\usepackage[colorlinks, citecolor=blue,linkcolor=blue]{hyperref}
\usepackage{textcomp}
\usepackage{subfigure}
\usepackage{graphicx}
\begin{document}
 \title{Engineering Entanglement Mechanically}
\author{Muhammad Asjad$^1$}
\author{Farhan Saif$^{2,*}$}
 \affiliation{$^1$Department of Electronics, Quaid-i-Azam University, 45320 \ Islamabad, Pakistan.\\
$^2$Centre for Advanced Mathematics and Physics, National University of Science and Technology, H12, Islamabad, Pakistan}
\begin{abstract}
\centerline{Abstract:}

We propose entanglement for hybrid optomechanical system consisting of Bose-Einstein condensate (BEC) inside a single-mode high-Q Fabry-Perot cavity with a vibrating end mirror (mechanical mirror). The intracavity field couples the vibrating end mirror with collective atomic density of the BEC. We show that the radiation pressure generates the stationary entanglement of three bipartite subsystems, \textit{i.e}, field-mechanical mirror, field-BEC and mechanical mirror-atoms. The resulting entanglement is fragile with respect to temperature.
\end{abstract} 
\keywords{mechanical action of light, mechanical mirror, bose-einstein condensation, entanglement, nanotechnology, nano-electro-mechanical systems\\
$^*$Tel. no.: +92 51 9064 2104, E-mail: fsaif@nust.camp.edu.pk}
\maketitle

Nano-optomechanical systems coupled with Bose-Einstein condensate provide an interesting play-ground both in theory and experiment. These systems are promising in quantum informatics \cite{schwab} and quantum metrology \cite{matrology}. Quantum correlations in these systems \cite{Gabriele} generates multipartite entanglement \cite{three}, thus these are suitable to study quantum teleportation \cite{prltelep}, quantum telecloning \cite{jmo} and entanglement swapping \cite{Pir06}. The most challenging goals of the modern experimental quantum mechanics is to create  multi-partite entangled states \cite{three}. Quantum entanglement has been widely studied in different systems such as optomechanical systems \cite{2,Vitali}, two atomic ensembles by sending pulses of coherent light through two atomic vapor cells \cite{3}, Bose-Einstein condensates trapped in double well \cite{4} and in optical lattice \cite{5}. In this paper we develop steady state off-resonant multipartite entanglement in hybrid system formed by Bose-Einstein condensate (BEC) inside a cavity with high-finesse, single mode optical cavity, and a moving end-mirror. The atom-light interaction is enhanced as the condensate atoms are collectively coupled to the same light mode. The intracavity field acts as a nonlinear spring which couples the BEC atoms with vibrating mirror of the cavity. We measure the entanglement between mechanical and atomic modes with intracavity field and also between mechanical and atomic mode themselves. The entanglement capabilities of optomechanical system are modified due to the back action induced by the atoms. We show that the entanglement is sensitive to temperature variation. Though bipartite entanglement, namely, atom-field and mirror-field entanglement survive at higher temperatures, however atom-mirror entanglement is relatively fragile and available at lower temperature scale. The use of mutual coupling between mechanical and atomic subsystems provides coherent quantum control at mesoscopic scale \cite{6}.
\paragraph*{} We consider a Bose-Einstein condensate (BEC) of N two-level atoms strongly interacting with a quantized single cavity mode of frequency, $\omega_\mathrm{c}$. The field inside the cavity forms one-dimensional optical lattice potential. The intracavity field is coupled to the external fields incident from one side  mirror with partial reflectivity, when the cavity is coherently driven by a laser light with frequency, $\omega_\mathrm{L}$, and amplitude, E. We consider the other end mirror, with 100\% reflectivity, of the optical cavity of length, L, as moving and following harmonic oscillation. The mirror oscillates with frequency, $\omega_m$, and in the absence of the radiation-pressure coupling it undergoes Brownian motion as it is connected with thermal environment. The system is open as cavity field is damped due to the leakage of photons through the fixed mirror, and the vibrating end mirror is connected to a bath, at finite temperature T. The Hamiltonian of the system made out of the intracavity field, the BEC, and the moving end mirror of the cavity is 
\begin{equation}
  \hat{H}=\hat{H}_{\mathrm{m}}+ \hat{H}_{\mathrm{a}}+\hat{H}_{\mathrm{T}}\,,\label{1}
\end{equation}
where, $\hat H_\mathrm{m}$ describes moving-end-mirror and its coupling to the light field, $\hat H_\mathrm{a}$ describes BEC and its coupling to the intracavity field and $\hat H_\mathrm{T}$ accounts for dissipation and coupling of subsystems to the thermal reservoirs.
\paragraph*{} The mirror-field Hamiltonian $H_\mathrm{m}$ is given explicitly \cite{cklaw} as, 
 \begin{equation}
\hat{ H}_\mathrm{m} = \hslash\, \Delta_{\mathrm{c}}\, \hat{c}^{\dag}\hat{c}\, + \dfrac{\hslash\,
\omega_{\mathrm{m}}}{2} (\hat{p}_{\mathrm{m}}^{2} + \hat{q}_\mathrm{m}^{2})-\hslash\, \zeta_{\mathrm{mc}} \,\hat{c}^{\dag}\hat{c}\,\hat{q}_\mathrm{m}-i\,\hslash\, E\,(\hat{c} - \hat{c}^{\dag})\,,\label{2}
 \end{equation}
where, $\Delta_\mathrm{c}=\omega_\mathrm{c}-\omega_\mathrm{L}$. Dimensionless momentum and position operators of the mechanical oscillator of mass \textit{m} are, respectively, $\hat{p}_{\mathrm{m}}$ and $\hat{q}_\mathrm{m}$, with commutation relation $[\hat{q}_\mathrm{m}, \hat{p}_{\mathrm{m}}]=i$. Furthermore, $\hat{\mathrm{c}}$ and $\hat{\mathrm{c}}^{\dag}$ are the annihilation and creation operators,  for the intracavity field which satisfy the commutation relation $[\hat{c}, \hat{c}^{\dag}]=1$. The input laser field of amplitude $|\mathrm{E}|=\sqrt{P\kappa/\hbar\omega_{\mathrm{L}}}$, populates the cavity mode which is coupled to the mechanical oscillator (vibrating mirror) with frequency $\omega_\mathrm{m}$ through radiation-pressure via the coupling parameter, $\zeta_{\mathrm{mc}}=\omega_c\sqrt{\hslash/m\omega_\mathrm{m}}/\mathrm{L}$. 
\paragraph*{} In order to describe the motion of the BEC atoms inside an optomechanical cavity, we assume that the atoms are trapped in one-dimensional optical lattice and their motion is quantized along the cavity axis. We also assume that the atom-field detuning $\Delta_\mathrm{a}=\omega_\mathrm{L}-\omega_\mathrm{a}$ is large, so that spontaneous emission is negligible, and we can adiabatically eliminate the internal excited state dynamics of the atoms. Our analysis is valid for weakly interacting BECs with little or no interactions, 
a situation that can be realized experimentally with Feshbach scattering resonances \cite{Vogels1997}. 
In addition, following the discussion in Ref.~\cite{ChoiNiu1999,ayub}, we consider that the analysis is also valid 
for strongly interacting homogeneous condensate, where, nonlinear term can be replaced by an effective 
potential provided the external modulation causes slight changes in density profile of the condensate.  In the presence of atom atom interaction, however, we employ Bose-Hubbard Hamiltonian~\cite{2b}. The atom-field Hamiltonian, $H_\mathrm{a}$, is written as \cite{petaviski},
\begin{equation}
\hat{H}_{\mathrm{a}}=\int\hat{\Psi}^\dag(x)\left[-\dfrac{\hslash^{2}}{2\,\mathrm{m}_{\mathrm{\mathrm{a}}}} \dfrac{d^{2}}{dx^{2}} +\hslash\, \dfrac{g^2(x)}{\Delta_\mathrm{a}}\hat{c}^{\dag}\hat{c} \right]\hat{\Psi}(x)\, dx\,,\label{3}
\end{equation}
where, $\hat{\Psi}(x)$ is the bosonic field annihilation operator for the atoms. The intracavity mode couples to the BEC atoms through the dipole interaction via $g(x)=g_o\cos(kx)$, where, \textit{k} is the wave number of the light field. The cavity dynamics of BEC can be described in a homogeneous two-mode model where, the macroscopically zero momentum state is only coupled to symmetric momentum states $\pm2\hslash k$ via absorption and stimulated emission of cavity photons \cite{7}. Accordingly, we can write the atomic field operator $\hat{\Psi}(x)$ as,
\begin{equation}
\hat{\Psi}(x) = [\hat{b}_{o} + \sqrt{2}\cos(2\,kx)\,\hat{b}_{2}]/\sqrt{L}\,,\label{4}
\end{equation}
where, $\hat{b}_\mathrm{o}$ and $\hat{b}_2$ being the bosonic annihilation operator of the corresponding modes. By inserting the ansatz into Eq.(\ref{3}), we get the following second quantized Hamiltonian operator in Bogoliubov approximation 
\begin{equation}
\hat{H}_\mathrm{a}= \dfrac{\hslash\, U_{\mathrm{o}}\mathrm{N}}{2}\,\hat{c}^{\dag}\hat{c} + \dfrac{\hslash\,\Omega}{2}(\hat{p}_\mathrm{a}^{2} + \hat{q}_\mathrm{a}^{2})+\zeta_{\mathrm{ac}}\,\hslash\, \hat{c}^{\dag}\hat{c}\,\hat{q}_\mathrm{a}\,,\label{5}
\end{equation} 
where, $\Omega= 4\omega_\mathrm{r} = 2\hslash\, k^2/m_\mathrm{a}$. The parameter $U_\mathrm{o}=g^2_\mathrm{o}/\Delta_\mathrm{a}$ is the optical lattice barrier depth per photon and represents the atomic back action on the field \cite{8}. Here $\zeta_{\mathrm{ac}}=\sqrt \mathrm{N} U_\mathrm{o}/2$, \textit{i.e.}, the single-atom interaction $U_\mathrm{o}$ is increased by the square root of number of the atoms, N. Along the cavity axis (x-axis), the intracavity field forms an optical lattice potential of period $\lambda/2$, and depth, $\hslash U_\mathrm{o}\hat{c}^{\dag}\hat{c}$. The Eq.(\ref{5}) describes a mechanical oscillator coupled to the cavity field through the radiation pressure, where  position and momentum operators, $\hat{q}_\mathrm{a}=(\hat{b}+\hat{b}^{\dag})/\sqrt2$ and $\hat{p}_\mathrm{a}=(\hat{b}-\hat{b}^{\dag})/i\sqrt2$, satisfy the commutation relation $[\hat{q}_\mathrm{a}, \hat{p}_\mathrm{a}]=i$. We call  this fictitious mirror as atomic-mirror, which is an analogy to the moving-end-mirror of the cavity.
\paragraph*{}For the full description of the hybrid optomecahnical system, we must include the effects of the dissipation, on the intracavity field, the damping of the mechanical oscillator, and the damping of the atomic-mirror. The Hamiltonian $H_\mathrm{T}$ in Eq.(\ref{1}) accounts for these processes, and these noise processes are included via standard quantum noise operators \cite{9}.
\paragraph*{} In order to describe the complete dynamics of the subsystems involved, an adequate choice is to use the formalism of the quantum Langevin equations. The explicit form of the set of  Langevin equations for the general system at hand reads as,
\begin{eqnarray}
\dot c &=&(-i\Delta_o +i\zeta_{\mathrm{mc}}\,q_\mathrm{m}-i\zeta _{\mathrm{ac}}\,q_\mathrm{a}-\kappa)c + E + \sqrt{2\kappa}\,c_{\mathrm{in}} \,, \nonumber \\
\dot q_\mathrm{m} &=&\omega_\mathrm{m}\, p_\mathrm{m}\,,\nonumber \\   
\dot p_\mathrm{m} &=&-\omega_{\mathrm{m}} q_\mathrm{m} + \zeta_{\mathrm{mc}}\, c^{\dag} c-\gamma_{\mathrm{m}}\, p_\mathrm{m} + f_{\mathrm{m}}\,,\label{6} \\ 
\dot q_\mathrm{a} &=&\Omega\, p_\mathrm{a}  \,,  \nonumber\\
\dot p_\mathrm{a} &=&-\Omega\, q_\mathrm{a} -\zeta_{\mathrm{ac}}\, c^{\dag}c\,, \nonumber
\end{eqnarray}  
where, $\Delta_\mathrm{o}=\Delta_\mathrm{c} + N\,U_\mathrm{o}/2 $. For simplicity we omit the \text{hat} sign from the operators in Eq.(\ref{6}) and in later calculations. Here, $\kappa$ and $\gamma_\mathrm{m} $ respectively characterize the dissipation of the cavity field, mechanical oscillator and collectively density excitations of the BEC, respectively. The cavity input noise is delta correlated in time domain, \textit{i.e}, $\langle c _{\mathrm{in}}(t)\,c^{\dag}_{\mathrm{in}}(t') \rangle =\delta (t-t')$ for $\hbar\omega_{\mathrm{c}}/K_\mathrm{B}\mathrm{T}>>1 $ and all other correlations are zero. Mechanical Brownian noise operator $f_{\mathrm{m}}$ with zero mean value is generally non-Markovian. However, the mechanical frequency never becomes larger then hundreds of MHz and even for cryogenic temperature, the correlation function of $f_{\mathrm{m}}$ can be approximated as $\langle f_\mathrm{m}(t)f_\mathrm{m}(t') \rangle =\gamma_\mathrm{m}\,(2n+1)\delta (t-t')$ \cite{10}, where,  $n=[\exp{\{\hslash\omega_\mathrm{m}/K_\mathrm{B}\mathrm{T}\}}-1]^{-1}$ is the equilibrium phonon number of the mechanical oscillator.
\paragraph*{} We now rewrite each Heisenberg operator of the Eq.(\ref{6}) as the sum of its steady state mean value and an fluctuation operator with zero mean value \textit{i.e} $q_\mathrm{m}=q_\mathrm{ms}+\delta q_\mathrm{m}$, $p_\mathrm{m}=p_\mathrm{ms}+\delta p_\mathrm{m}$, $q_\mathrm{a}=q_\mathrm{as}+\delta q_\mathrm{a}$, $p_\mathrm{a}=p_\mathrm{as}+\delta p_\mathrm{a}$, $c=c_\mathrm{s}+\delta c$. Neglecting the atomic losses due to heating we obtain the following linearized Heisenberg-Langevin equations 
\begin{eqnarray}
\delta\dot q_\mathrm{m} &=& \omega_\mathrm{m}\, \delta p_\mathrm{m}\,, \nonumber\\
\delta\dot p_\mathrm{m}&=&-\omega_\mathrm{m}\, \delta q_\mathrm{m} + \chi_{\mathrm{mc}}\, \delta X-\gamma_\mathrm{m}\,\delta p_\mathrm{m}+f_\mathrm{m}\,, \nonumber \\  
\delta\dot q_\mathrm{a}&=&\Omega \,\delta p_\mathrm{a}\,,\label{7} \\
\delta\dot p_\mathrm{a}&=&-\Omega\, \delta q_\mathrm{a} -\chi_{\mathrm{ac}}\, \delta X\,, \nonumber \\
\delta\dot X&=& \Delta\, \delta Y-\kappa \,\delta X+\sqrt {2\kappa}\,X_{\mathrm{in}}\,, \nonumber\\
\delta\dot Y&=&-\Delta\, \delta X+\chi_{\mathrm{mc}}\,\delta q_m -\chi_{\mathrm{ac}}\,\delta Q-\kappa\,\delta Y+\sqrt{2\kappa}\,Y_{\mathrm{in}}\,,\nonumber
\end{eqnarray} 
where, $\Delta=\Delta_\mathrm{o}-|c_\mathrm{s}|^2\left(\zeta^2_{\mathrm{mc}}/\omega_\mathrm{m} - \zeta^2_{\mathrm{ac}}/\Omega\right)$ is the effective cavity detuning . Here we consider the phase of the cavity field so that $c_\mathrm{s}$ is real and positive, and we have defined the cavity field quadratures $\delta X=(\delta c+\delta c^{\dag})/\sqrt{2}$ and  $\delta Y=(\delta c-\delta c^{\dag})/i\sqrt{2}$, and the corresponding Hermitian input noise operators  $\delta X_{\mathrm{in}}=(\delta c_{\mathrm{in}}+\delta c^{\dag}_{\mathrm{in}})/\sqrt{2}$ and $ \delta Y_{\mathrm{in}}=(\delta c-\delta c^{\dag}_{\mathrm{in}})/i\sqrt{2} $. The linearized quantum Langevin equations show that the fluctuations of mechanical mirror and atomic-mirror are now coupled to the cavity field quadrature fluctuations by the effective couplings $\chi_{\mathrm{mc}}=\zeta_{\mathrm{mc}}\,c_\mathrm{s}\sqrt{2}$ and $\chi_{\mathrm{ac}}=\zeta_{\mathrm{ac}}\,c_\mathrm{s}\sqrt{2}$, which can be made very large by increasing the amplitude $c_\mathrm{s}$ of the intracavity field. Moreover, significant entanglement of optical field with mechanical oscillator and atomic mirror become possible.
\paragraph*{} The system of linearized quantum Langevin Eqs.(\ref{7}) can be written in compact matrix form as $\dot R(t)=M\,R(t)+F(t) $, where, $R$ is the vector of the quadrature fluctuations and $F$ correspond to noises. Here $M$ is drift matrix. Since the quantum noises are white in nature and the dynamics is linearized, the state of the system will be continuous variable (CV) tripartite Gaussian state, and is completely determined by first and second moments. The equations of motions for first and second moments are
\begin{figure*}
    \centering
 \subfigure[]{\includegraphics[scale=.21]{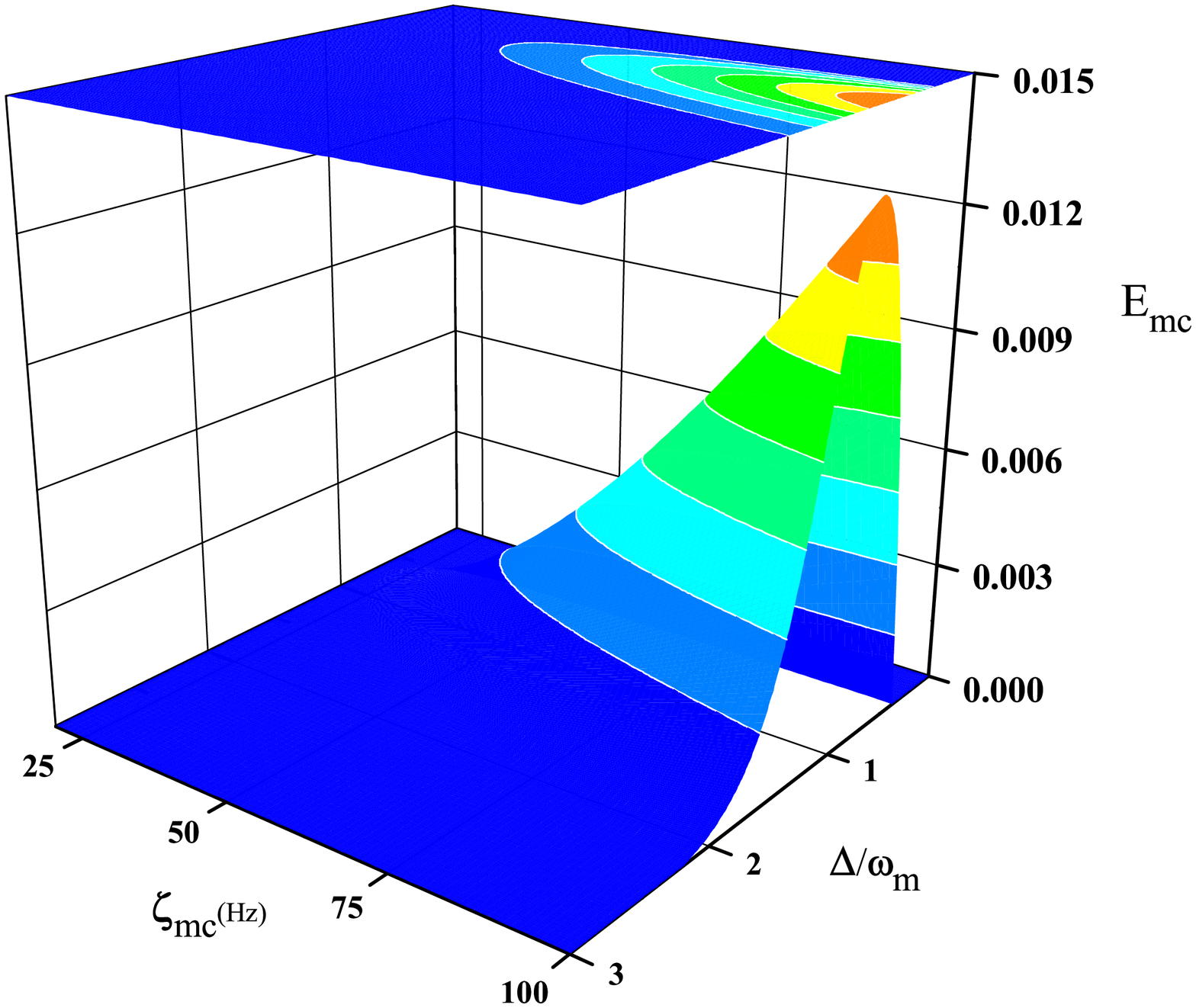}\label{emf}}
 \subfigure[]{\includegraphics[scale=.21]{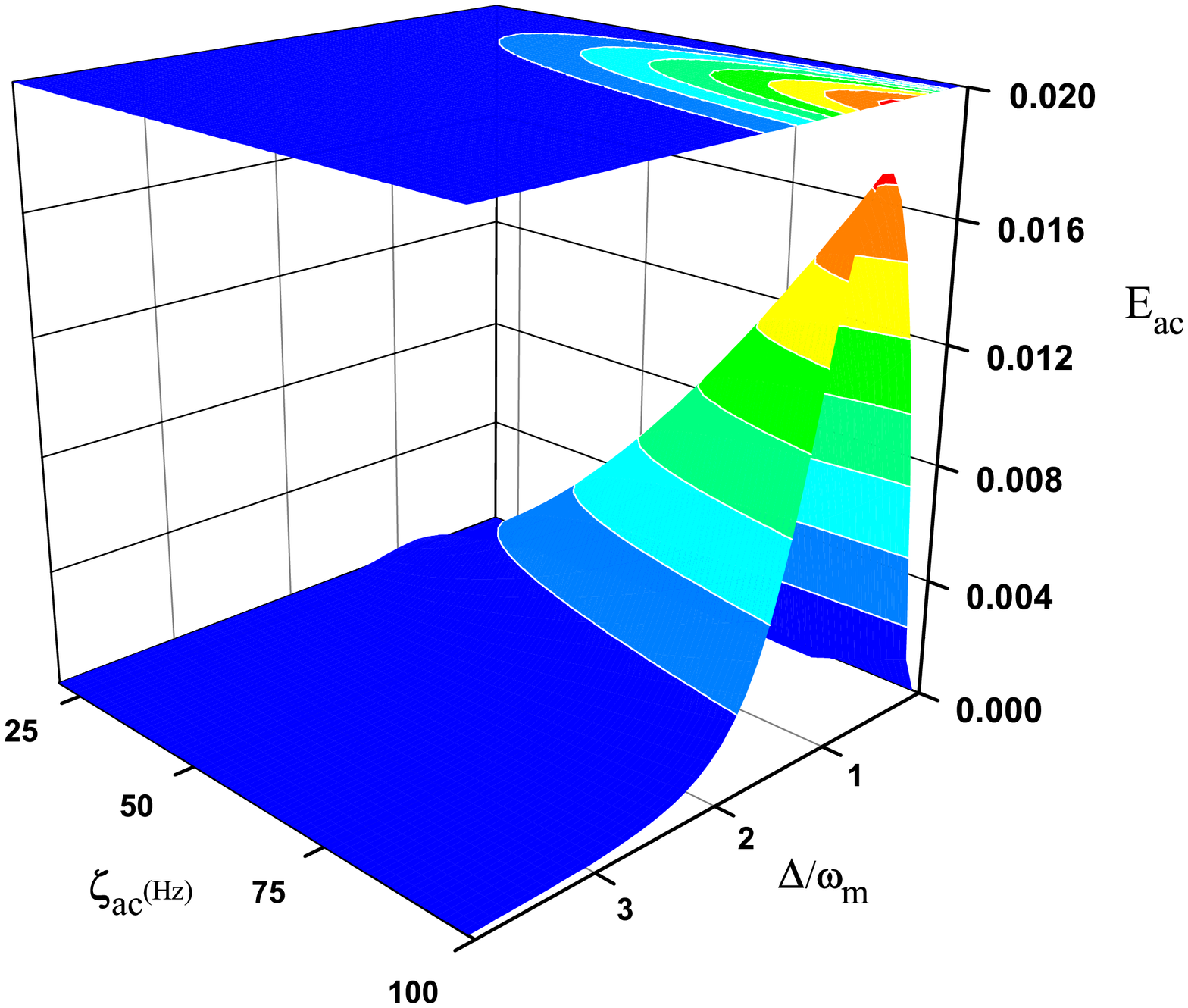}\label{eaf}}
 \subfigure[]{\includegraphics[scale=.23]{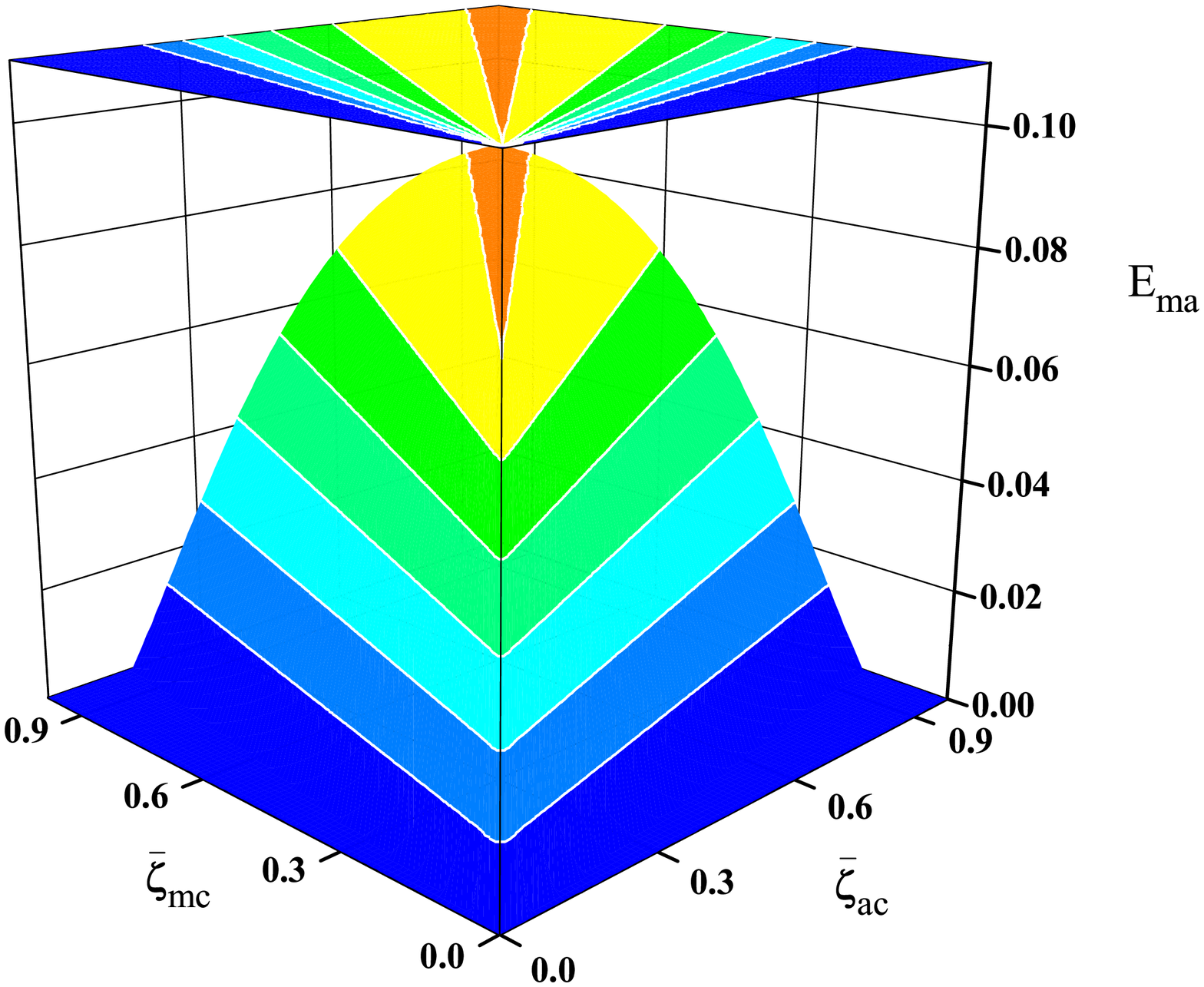}\label{eam}}
\caption{(Color online) (a) Plot of logarithmic negativity $E_{\mathrm{mc}}$ as a function of the normalized detuning $\Delta/\omega_m$ and coupling of mechanical mirror with optical field $\zeta_{\mathrm{mc}}$ for $\zeta_{\mathrm{ac}}=0.7\,\zeta_{\mathrm{mc}}$ and $\omega_m \simeq \Omega$. The optical cavity length $\mathrm{L} = 1\, \mathrm{mm}$, driven by a laser with wavelength $\lambda =\, 1000\, \mathrm{nm}$, power $P = 50\,\mathrm{mW}$ and $\omega_m \simeq \Omega$. The mechanical mirror has a frequency $\omega_\mathrm{m}/2\pi=10\,\mathrm{MHz}$, and damping rate $\gamma_{\mathrm{m}}/2\pi=100 \, \mathrm{Hz}$, its temperature is $\mathrm{T}=100\, \mathrm{mK}$ and cavity finesse is $F=1.07 \times 10^4$.(b) $E_{\mathrm{ac}}$ against $\Delta/\omega_m$ and  $\zeta_{\mathrm{ac}}$ for $\zeta_{\mathrm{mc}} = \mathrm{i}\,\zeta=\mathrm{i}\,\omega_c\sqrt{\hbar/m\omega_m}/\mathrm{L}$ ($\mathrm{i}=0.01$) and $\omega_\mathrm{m}= \Omega$. (c) $E_{\mathrm{ma}}$ against normalized coupling of mechanical mirror $\bar{\zeta}_\mathrm{mc}=\zeta_{\mathrm{mc}}/\omega_{\mathrm{m}}$ and atomic mirror $\bar{\zeta}_\mathrm{ac}=\zeta_\mathrm{ac}/\omega_{\mathrm{m}}$ for $\Delta = 0.6 \, \omega_\mathrm{m}$ and temperature is $\mathrm{T}=1\,\mathrm{\mu K}$.} 
\label{ent}
\end{figure*}
\begin{eqnarray}
\dot d&=&M d\,, \nonumber\\
\dot{V}&=&M V + V M^t + V_\mathrm{F}\,, \label{8}
\end{eqnarray}
where, $d=\langle R\rangle$ is the displacement vector and $V_{ij}=\langle (R_i\,R_j+R_j\,R_i)-2d_i\,d_j \rangle$ is $6\times6$ covariance matrix. Here $V_\mathrm{F}$ is the covariance matrix of noises and in superscript $t$ describes transpose of matrix. The system will be stable and reaches its steady state only if all the eigenvalues of the drift matrix $M$ are in the left half plane. The stability conditions can be obtained by applying Routh-Hurwitz criterion \cite{11}. In steady state covariance matrix fulfills the Lyapunove equation,
\begin{equation}
M V + V M^t=-V_\mathrm{F}. \label{9}
\end{equation}
which is the linear matrix equation and can be straight forwardly solved however the general exact expression is too cumbersome and is not reported here. One can extract all the information about the steady state of the system from the correlation matrix. The entanglement between any two of the three bipartite states can be measured by tracing out one of the three modes, that is, mechanical, atomic or optical mode. 
\paragraph*{}We can quantify the steady state entanglement by considering the logarithmic negativity, $E_N$ \cite{12}. As $E_N$ only measures the entanglement of bipartite system, we measure the entanglement of atomic-mirror and moving end mirror separately with optical field, in addition we also measure the entanglement between atomic-mirror and moving end mirror. We denote the logarithmic negativities for the mechanical mirror-field, atomic mirror-field and mechanical mirror atomic mirror bimodal partition as $E_{\mathrm{mc}}$, $E_{\mathrm{ac}}$ and $E_{\mathrm{ma}}$, respectively.
\paragraph*{} At first we measure the entanglement between mechanical mirror and optical field $E_\mathrm{mc}$, which is obtained by tracing out the atomic-mirror mode, \textit{i.e}, removing the rows and columns of $V$ correspond to atomic-mirror. The reduced state is still Gaussian and fully characterized by $4\times4$ matrix $V_{\mathrm{mc}}$ and is given by
\begin{figure*}
    \centering
 \subfigure[]{\includegraphics[scale=.21]{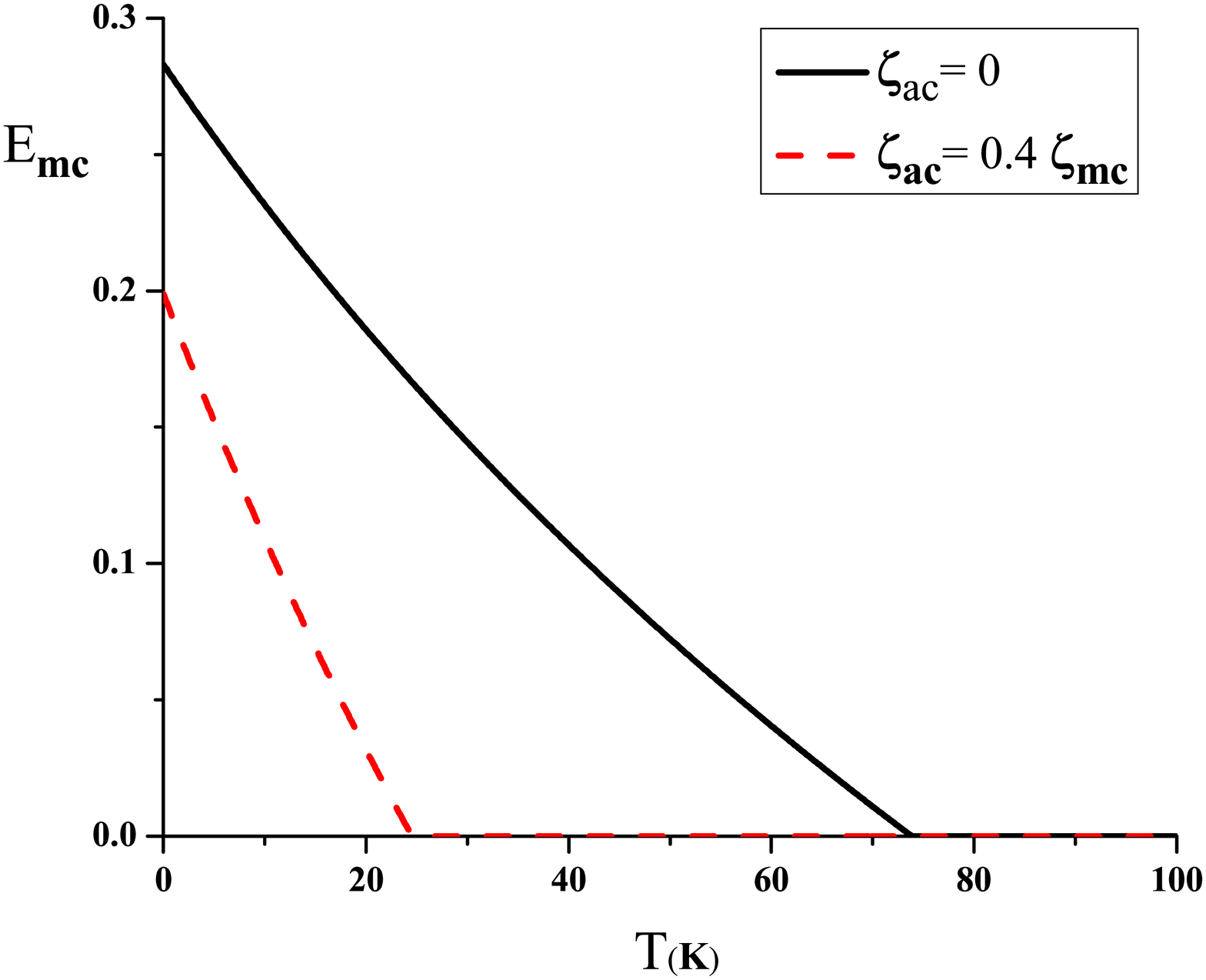}\label{emt}}
 \subfigure[]{\includegraphics[scale=.21]{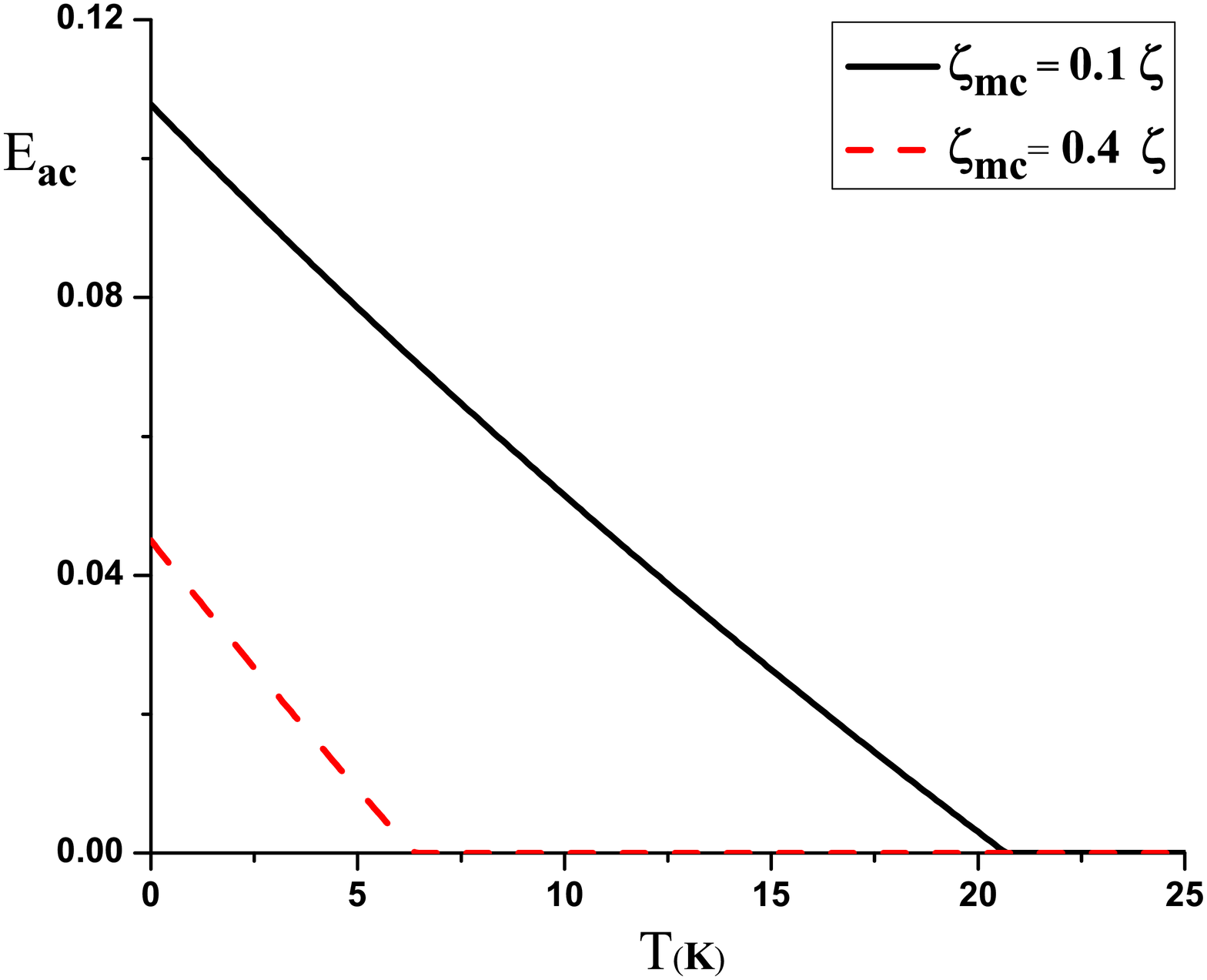}\label{eat}}
 \subfigure[]{\includegraphics[scale=.21]{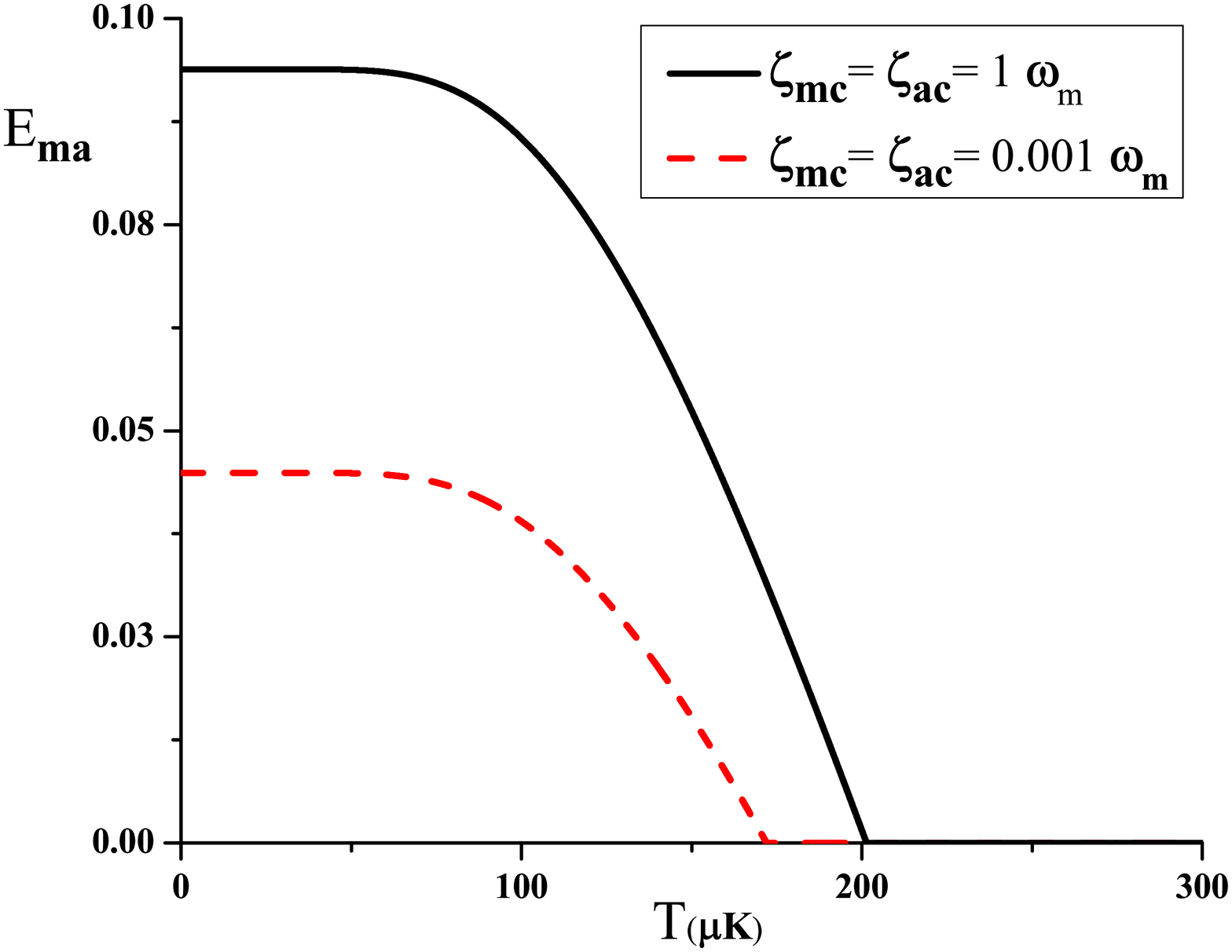}\label{eamt}}
\caption{(Color online) (a) Plot of logarithmic negativity $E_{\mathrm{mc}}$ as a function of environmental temperature T for $\Delta\simeq\,0.5\,\omega_{\mathrm{m}}$ and $\omega_\mathrm{m} \simeq \Omega$. The black line (solid) for $\zeta_{\mathrm{ac}}= 0$ and the red line (dashed) for $\zeta_{\mathrm{ac}}=0.4\,\zeta_{\mathrm{mc}}$.(b) $E_{\mathrm{ac}}$ against the environmental temperature T for $\Delta\simeq0.5\,\omega_{\mathrm{m}}$, $\zeta_{\mathrm{ac}}=100\,\mathrm{Hz}$ and $\omega_m \simeq \Omega$. $\zeta_{\mathrm{mc}} = \mathrm{i}\,\zeta=\mathrm{i}\,\omega_c\sqrt{\hbar/m\omega_m}/\mathrm{L}$, the black (solid) line for $\mathrm{i}=0$ and red dashed line (dashed) for $\mathrm{i}=0.4$  (c) $E_{\mathrm{ma}}$ against temperature T for $\Delta\simeq0.6\,\omega_{\mathrm{m}}$ and $\omega_m \simeq \Omega$. The black (solid) and red dashed (dashed) lines for $\zeta_{\mathrm{mc}}=\zeta_{\mathrm{ac}}=\omega_{\mathrm{m}}$ and $\zeta_{\mathrm{mc}}=\zeta_{\mathrm{ac}}=0.001\,\omega_{\mathrm{m}}$ respectively. The other parameters are the same as in Fig.\ref{ent}. } 
\label{entt}
\end{figure*}
\begin{center}
$V_{\mathrm{mc}}$=\(
 \begin{bmatrix}
\mathcal{X} & \mathcal{Z} \\ \mathcal{Z}^t & \mathcal{Y}
 \end{bmatrix}
\).
\end{center} 
Here, $\mathcal{X}$, $\mathcal{Y}$, $\mathcal{Z}$ are $2 \times 2$ matrices, where matrix $\mathcal{X}$ corresponds to the mechanical mirror, $\mathcal{Y}$ refer to the intracavity mode and $\mathcal{Z}$  describes the correlation between these two systems.  In order to measure the entanglement between mechanical mirror and optical field, we consider the logarithmic negativity $E_{\mathrm{mc}}$, which, in case of CV, $E_{\mathrm{mc}}$ can be defined as \cite{Adesso} 
\begin{equation}
E_{\mathrm{mc}}=\mathrm{max}[0,-\ln 2\,\varepsilon]\,,\label{55}
\end{equation}
 where, $\varepsilon\equiv 2^{-1/2}\{\sum(V_{\mathrm{mc}})-[\sum(V_{\mathrm{mc}})^2-4\det V_{\mathrm{mc}}]^{1/2}\}^{1/2}$, with $\sum(V_{\mathrm{mc}})\equiv \det \mathcal{X} + \det\mathcal{Y}-2\det\mathcal{Z}$. The Gaussian state gets entangled only if the eigenvalue $\varepsilon<1/2$, and it is  Simon's necessary and sufficient entanglement non-positive partial transpose criterion of the Gaussian states \cite{Simon}. This condition can also be written as $4\det V_{\mathrm{mc}}<\sum(V_{\mathrm{mc}})-1/4$.
 \paragraph*{} For our numerical calculation we take the parameters from the experimental work reported in \cite{6,7,15}. In Fig.\ref{ent} we plot the steady state entanglement between mechanical mirror and optical field Fig.\ref{emf}, atomic mirror and optical field Fig.\ref{eaf} and mechanical and atomic mirrors Fig.\ref{eam} measured by the logarithmic negativities $E_{\mathrm{mc}}$, $E_{\mathrm{ac}}$ and $E_{\mathrm{ma}}$, respectively. In Fig.\ref{emf} we plot the $E_{\mathrm{mc}}$ against normalized detuning $\Delta/\omega_{\mathrm m}$ and coupling rate of the mechanical mirror with cavity field $\zeta_{\mathrm{mc}}$ for $\zeta_{\mathrm{ac}}=0.7\,\zeta_{\mathrm{mc}}$ and $\omega_{\mathrm{m}}\simeq\Omega$. The entanglement measure between mechanical mirror and optical field increases with $\zeta_{\mathrm{mc}}$ and is present only for small interval of values of $\Delta$ around $\Delta\simeq 0.5\,\omega_\mathrm{m}$.
\paragraph*{} Similarly, the steady state entanglement of the atomic mirror and intracavity field is measured by the logarithmic negativity $E_{\mathrm{ac}}$ as a function of the normalized detuning and the coupling of the atomic mode with intracavity field for $\zeta=0.01\,\zeta_{\mathrm{mc}}$ and $\omega_{\mathrm{m}}\simeq\Omega$, as shown in Fig.\ref{eaf}. The features observed in the entanglement measure between atomic and cavity modes are similar to those present in $E_{\mathrm{mc}}$. The Entanglement measure between mechanical mirror and optical field increases with $\zeta_{\mathrm{ac}}$ and present only for small interval of values of $\Delta$ around $\Delta\simeq 0.5\,\omega_\mathrm{m}$.
\paragraph*{} Analogously, we measure the entanglement between mechanical mirror and atomic mirror. We numerically calculate the logarithmic negativity, $E_{\mathrm{ma}}$, associated with the steady state correlation matrix formed by the atomic and mechanical modes, by tracing out the cavity mode shown in Fig.\ref{eam}. Here the logarithmic negativity, $E_{\mathrm{ma}}$, is calculated as a function of the normalized coupling of the mechanical $\zeta_{\mathrm{mc}}/\omega_{\mathrm m}$ and atomic mirrors with intracavity field $\zeta_{\mathrm{ac}}/\omega_{\mathrm m}$ for $\Delta\simeq0.6\,\omega_{\mathrm{m}}$ and $\omega_{\mathrm{m}}\simeq\Omega$.
\paragraph*{} It is important to understand the behavior of entanglement with respect to the temperature T. In Fig.\ref{entt} we show the logarithmic negativity $E_{i}$, where, ($i$ = mc, ac, ma) as a function of the environmental temperature T. In Fig.\ref{emt}, it is noted that the entanglement between mechanical mirror and intracavity field is very robust with respect to temperature. The black solid line for $\zeta_{\mathrm{ac}}=0$ shows a significant amount of entanglement is present upto temperature $60\,\mathrm{K}$. The red dashed line for $\zeta_{\mathrm{ac}}= 0.4\,\zeta_{\mathrm{mc}}$ in Fig.\ref{emt} shows that the entanglement between mechanical mirror and intracavity mode persist only upto $20\, \mathrm{K}$. Similarly, the robustness of the entanglement between atomic mirror and intracavity mode as a function of the reservoir temperature of the mechanical mirror is shown in Fig.\ref{eat}. The black solid and red dashed lines are for $\zeta_\mathrm{mc}=0.1\,\zeta$ and $\zeta_\mathrm{mc}=0.4\,\zeta$, respectively. The BEC mode absorbs some phonons taken from the mechanical mirror by the field and its temperature is increased. There is no direct connection between BEC and thermal environment for zero interaction between mechanical mirror and cavity field, hence $E_{ac}$ is independent of temperature as the interaction is zero, \textit{i.e}, $\zeta_{\mathrm{mc}}=0$. In Fig.\ref{eat}, the black solid line for a smaller value  $\zeta_{\mathrm{mc}}=0.1\,\zeta$, of the interaction between mechanical mirror and intracavity field, shows that $E_{\mathrm{ac}}$ persists upto $20\, \mathrm{K}$. The red dashed line in Fig.\ref{eat} shows the entanglement between atomic and cavity field modes only presents upto $5\,\mathrm{K}$ as the coupling between mechanical mirror and intracavity is increased, that is, $\zeta_{\mathrm{mc}}=0.4\,\zeta$. The steady state entanglement between mechanical mode and atomic mode as a function of the temperature is shown in Fig.\ref{eamt} for two different values of the coupling of the cavity field with atomic and mechanical modes. The $E_{\mathrm{ma}}$ is fragile with respect to temperature. The black solid line and red dashed line are for $\zeta_{\mathrm{mc}}=\zeta_{\mathrm{ac}}=\omega_{\mathrm{m}}$ and  $\zeta_{\mathrm{mc}}=\zeta_{\mathrm{ac}}=0.001\,\omega_{\mathrm{m}}$, respectively.
 \begin{figure}[tb]
\centering
\includegraphics[scale=.32]{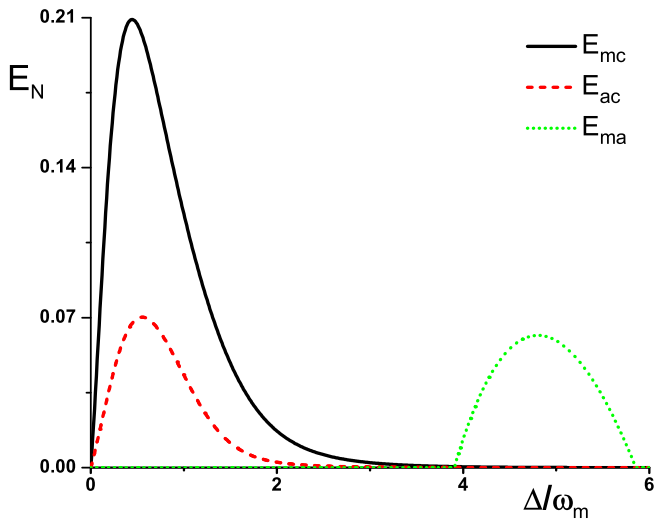}
\caption{(color online) Plot of Logarithmic negativity $\mathrm{E_N}$ as function of normalized detuning $\Delta/\omega_{\mathrm m}$ for  $\omega_{\mathrm m}=10 \,\mathrm{MHz}$ and $\Omega=0.9\,\omega_{\mathrm m}$. Mirror-field entanglement in the absence of BEC mode (black solid line) and BEC-field entanglement in the absence of mechanical mode (red dashed line). Moreover, the entanglement between mechanical mode and BEC mode in the absence of cavity mode (green dotted line) for $\Omega=\omega_{\mathrm m}=2\pi\times1\,\mathrm{MHz}$. Furthermore, $\zeta_{\mathrm{ac}}=200\,\mathrm{Hz}$ and $\zeta_{\mathrm{mc}}=300\,\mathrm{Hz}$ and temperature $\mathrm{T}=1\,\mathrm{\mu K}$. The other parameters are the same as in Fig.\ref{ent}.} \label{tri}
 \end{figure}
\paragraph*{} In Fig.\ref{tri} we plot the steady state entanglement of mechanical mirror with intracavity field (black line) versus normalized detuning $\Delta/\omega_{\mathrm m}$ and compere it with, steady state entanglement of the BEC mode with intracavity mode (red dashed line) and the steady state entanglement of the BEC mode and mechanical mode (green dotted line). Fig.\ref{tri} shows that the mirror-field and atom-field entanglement exist simultaneously, whereas atom-mirror entanglement is observed only for large values of the detuning.            
\paragraph*{} For experimental realization of the generated entanglement, one has to measure several quadrature correlations \cite{Duan} and these have been experimentally measured for the entanglement of the two optical modes \cite{16}. However, in our case we consider another Fabry Perot cavity adjacent to first one and is driven by a weak laser field, a scheme of this kind has been discussed in  \cite{Vitali}, in which the position and momentum of the mechanical mirror is directly measured by homodyning the out put of the second cavity for suitable values of the detuning and decay rate of the second cavity whereas the quadratures are directly measured by homodyning out put of the cavity. Finally, the quadratures of the BEC mode are measured by homomodning a weak field fed into the cavity as in Ref. \cite{esteve}.         
\paragraph*{} In conclusion, we have studied the hybrid quantum correlation of tripartite system which consists of optical cavity, a Bose-Einstein condensate mode, and a mechanical mirror mode of the Fabry-Perot cavity. We have shown steady state entanglement of the three bipartite subsystems: the mechanical mirror-field, the atomic mirror-field, and the mechanical mirror-atomic mirror by quantifying in term of logarithmic negativity. The resulting entanglement is fragile with respect to temperature. The realization of such systems opens new horizons for the realization of the quantum memories \cite{memory} and interfaces for the quantum information processing.                                  


\begin{thebibliography}{99}
\bibitem{schwab} K. C. Schwab and M. L. Roukes, phys. Today \textbf{58}, 36 (2005)
\bibitem{matrology}  M. Napolitano, M. Koschorreck, B. Dubost, N. Behbood, R. J. Sewell and M. W. Mitchell,    Nature \textbf{471}, 486  (2011).
\bibitem{Gabriele} Gabriele De Chiara, Mauro Paternostro and G. Massimo Palma, Phys. Rev. A \textbf{83}, 052324 (2011). 
\bibitem{three} A. S. Coelho, F. A. S. Barbosa, F. A. S. Barbosa, K. N. Cassemiro, A. S. Villar, M. Martinelli, and P. Nussenzveig, Science \textbf{326}, 823 (2009). 
\bibitem{prltelep} S. Mancini, D. Vitali, and P. Tombesi, Phys. Rev. Lett.
\textbf{90}, 137901 (2003). 
\bibitem{jmo} S. Pirandola S. Mancini, D. Vitali, and P. Tombesi, J. Mod.
Opt. \textbf{51}, 901 (2004).
\bibitem{Pir06} S. Pirandola, D. Vitali, P. Tombesi, and S. Lloyd, Phys.
Rev. Lett. \textbf{97}, 150403 (2006).
\bibitem{2}  S. Mancini, V. Giovannetti, D. Vitali and P. Tombesi, Phys. Rev. Lett. \textbf{88}, 120401 (2002); Jing Zhang, Kunchi Peng, and Samuel L. Braunstein, Phys. Rev. A \textbf{68}, 013808 (2003).
\bibitem{Vitali} D. Vitali, S. Gigan, A. Ferreira, H. R. B\"{o}hm, P. Tombesi, A. Guerreiro, V. Vedral, A. Zeilinger, and M. Aspelmeyer, Phys. Rev. Lett. \textbf{98}, 030405 (2007);
\bibitem{3} B. Julsgaard Alexander Kozhekin and Eugene S. Polzik, Nature (London) \textbf{413}, 400 (2001)
\bibitem{4} J. Chen et al., Phys. Lett. A \textbf{360}, 429 (2007).
\bibitem{5} L. You, Phys. Rev. Lett. \textbf{90}, 030402 (2003).
\bibitem{6} M. Paternostro, G. De Chiara and G.M. Palma, Phys. Rev. Lett. \textbf{104}, 243602 (2010).
\bibitem{cklaw}  C. K. Law, Phys. Rev. A \textbf{51}, 2537 (1995).
\bibitem{Vogels1997} J. M. Vogels et al., Phys. Rev. A 56, R1067 (1997); 
Ph. Courteille et al Phys. Rev. Lett. 81 69 1998.
\bibitem{ChoiNiu1999}D. I. Choi, and Q. Niu, Phys. Rev. Lett. 82, 2022 1999.
\bibitem{ayub} M. Ayub, and F. Saif, Phys. Rev. A {\bf 85}, 023634 (2012).
\bibitem{2b} M. Asjad, and F. Saif, Proceedings of 'Conference on Applications and Methods of Physics' Islamabad, Pakistan, pp. 12-15, Ed. A. Qadir, R. Khalid, and F. Saif.  
\bibitem{petaviski} L. Pitaevskii and S. Stringari, \textit{Bose-Einstein condensation} (Oxford University Press, Oxford , 2003).
\bibitem{7} F. Brennecke, S. Ritter, T. Donner, T. Esslinger, Science \textbf{322}, 235 (2008).
\bibitem{8} C. Maschler and H. Ritsch, Phys. Rev. Lett. \textbf{95}, 260401 (2005).
\bibitem{9}  C. W. Gardiner, Quantum Noise (Berlin: Springer 1991).
\bibitem{10} C. Genes, D. Vitali, and P. Tombesi, Phys. Rev. Lett. \textbf{77}, 050307 (2008).
\bibitem{11} E. X. DeJesus and C. Kaufman, Phys. Rev. A \textbf{35}, 5288 (1987).
\bibitem{12} G. Vidal and R.F Warner, Phys. Rev. A \textbf{65}, 032314 (2002).
\bibitem{Adesso} G. Adesso, Alessio Serafini, and Fabrizio Illuminati, Phys. Rev. A \textbf{70}, 022318 (2004).
\bibitem{Simon} R. Simon, Phys. Rev. Lett. \textbf{84}, 2726 (2000).
\bibitem{15} D. Kleckner, William Marshall, Michiel J. A. de Dood, Khodadad Nima Dinyari, Bart-Jan Pors, William T. M. Irvine, and Dirk Bouwmeester, Phys. Rev. Lett. \textbf{96}, 173901 (2006); S. Gigan, H. B\"{o}hm, M. Paternostro, F. Blaser, G. Langer, J. Hertzberg, K. Schwab, D. Bäuerle, M. Aspelmeyer, and A. Zeilinger, Nature  (London) \textbf{444}, 67 (2006); O. Arcizet, P.-F. Cohadon, T. Briant, M. Pinard, and A. Heidmann, ibid. \textbf444, 71 (2006); D. Kleckner and D. Bouwmeester, ibid, \textbf{444}, 75 (2006); H. R. B\"{o}hm, S. Gigan, F. Blaser, A. Zeilinger, and M. Aspelmeyer, Appl. Phys. Lett. \textbf{89}, 223101 (2006).
\bibitem{Duan} Lu-Ming Duan, G. Giedke, J. I. Cirac, and P. Zoller,  Phys. Rev. Lett. \textbf{84}, 2722 (2000).
\bibitem{16} Julien Laurat, Gaelle Keller, Jose Augusto Oliveira-Huguenin, Claude Fabre, Thomas Coudreau, Alessio Serafini, Gerardo Adesso, and Fabrizio Illuminati, J. Opt. B \textbf {7}, S577 (2005).
\bibitem{esteve} J. Estève, C. Gross, A. Weller, S. Giovanazzi, and M. K. Oberthaler,  Nature (London) \textbf{455}, 1216 (2008); M. Greiner, O. Mandel, T. Esslinger, T.W. Hänsch and I. Bloch, ibid. \textbf{415}, 39 (2002); F. Brennecke T. Donner, S. Ritter, T. Bourdel, M. Köhl, and T. Esslinger, ibid. \textbf{450}, 268 (2007); Y. Colombe, T. Steinmetz, G. Dubois, F. Linke, D. Hunger, and J. Reichel, ibid. \textbf{450}, 272
    (2007).
\bibitem{memory} Morgan P. Hedges, Jevon J. Longdell, Yongmin Li, and Matthew J. Sellars, Nature  (London) \textbf{465}, 1052 (2010).
\end{thebibliography}
\end{document}